\documentclass{aa}
\usepackage{graphicx}
\begin{document}

%   \thesaurus{07.13.2 ; 05.03.1 ; 02.09.1}    
   \title{Effective stability of the Trojan asteroids}

   \author{Ch. Skokos
          \inst{1}
          \and
          A. Dokoumetzidis\inst{2}
          }

   \offprints{Ch. Skokos (hskokos@cc.uoa.gr)}

   \institute{Research Center for Astronomy, Academy of Athens,
              14 Anagnostopoulou Str., GR-106 73, Athens, Greece
         \and
             School of Pharmacy, University of Athens,
             Panepistimiopolis, GR-157 71, Zografos, Athens, Greece \\
             }

   \date{Received 19 October 2000; accepted 7 December 2000}

   \abstract{We study the spatial circular restricted problem 
of three bodies in the light
of Nekhoroshev theory of stability over  large time intervals. We
consider in particular the Sun-Jupiter model and the Trojan asteroids in the
neighborhood of the Lagrangian point $L_4$. We find a region of
effective stability around the point $L_4$ such that if the initial
point of an orbit is inside this region the orbit is confined in a slightly
larger neighborhood of the equilibrium (in phase space) for a very long time
interval. By combining analytical methods and numerical approximations we are
able to prove that stability over the age of the universe is guaranteed on a
realistic region, big enough to include one real asteroid. By comparing this 
result with the one obtained for the planar problem we see that the regions of
stability in the two cases are of the same magnitude.      
\keywords{minor planets, asteroids -- celestial
mechanics -- instabilities}}
\maketitle
%
%________________________________________________________________

\section{Introduction}

The study of a Hamiltonian system in the neighborhood of an elliptic
equilibrium point is of interest in many fields of mathematical physics and
astronomy. Let us consider an analytic Hamiltonian $H$ with $n$ degrees of
freedom, having an elliptic equilibrium point. Rigorous results
proving the existence of orbits which do not leave a neighborhood of the
equilibrium can be given in the frame of KAM theory, under generic
conditions of non-resonance and non-degeneracy. KAM guarantees the existence
of many \mbox{$n$--dimensional} invariant tori around the equilibrium point.
However, such invariant tori do not fill an open region, i.e. the possibility
of the so--called Arnold diffusion cannot be excluded, except for the two
dimensional case. 

An alternative approach is to look for results which are valid over a
finite time interval, but give an effective bound on the Arnold diffusion.
This goal can be achieved by constructing the normal form of the Hamiltonian
around the elliptic equilibrium point. Normal forms are a standard tool in
Celestial Mechanics for studying the dynamics in the neighborhood of an
elliptic equilibrium point. Usually these normal forms are obtained as
divergent series but their truncation makes them useful. Roughly
speaking one shows that the system admits a number of approximate integrals,
whose time  variation  can be controlled to be small for an
extremely long time.  In these cases we have effective stability, i.e. even
when an  orbit is not stable, the time needed for it to leave the neighborhood
of the equilibrium is larger than the expected lifetime of the physical
system studied. This is the basis to
derive the classical Nekhoroshev's estimates (Nekhoroshev \cite{nek1}).
Different proofs of the  Nekhoroshev theorem were given by
Benettin et al. (\cite{benetal}), Benettin \& Gallavotti (\cite{bengal}),
Giorgilli \& Zehnder (\cite{giorz}) and P\"{o}schel (\cite{posc}).

A scientific field where the Nekhoroshev theory has been applied is the
problem of the stability of the Trojan asteroids. In recent years this
problem has been investigated by a number of researchers, both numerically and
analytically. The numerical investigations deal with  the evolution in time
of a sample of orbits, in sophisticated realistic models of the solar system,
and the statistical study of these orbits (Milani \cite{mil1}, \cite{mil2};
Levison et al. \cite{levi}; Tsiganis et al. \cite{tsig}).

In analytical studies simpler models of the system have been used such as the
two dimensional (2D) planar, and the three dimensional (3D) spatial restricted
three body problem. According to Nekhoroshev
theory, one has to estimate the rate of diffusion around the elliptic
equilibrium Lagrangian point $L_4$. Because of the symmetries of the system
the same study is valid for the $L_5$ point. The problem has been previously
investigated by Giorgilli et al. (\cite{gioretal}), Celletti \& Giorgilli
(\cite{celgior}) and Giorgilli \& Skokos (\cite{giorskok}) (hereafter Paper
I). Also analytical stability results for particular orbits were provided by
Celletti \& Ferrara  (\cite{celfer}) who studied the orbit of the asteroid
Ceres and by Jorba \& Villanueva  (\cite{jorvil}) who studied a stable
periodic orbit around the equilibrium point $L_5$.

The estimation of the region of effective stability by Giorgilli et al.
(\cite{gioretal}) and Celletti \& Giorgilli (\cite{celgior}) was realistic
but the region where the real asteroids were actually found was larger by a
factor 300 (in the best case) to 3,000, compared to the theoretical stability
region. The theoretical estimation was significantly improved in Paper I, since
the region found in the planar restricted three body problem was big enough to
include 4 real asteroids, while most of them failed to be inside this region by
a factor only 10. This  result underlined the fact that
Nekhoroshev  theory can give meaningful estimates, applicable to real
systems. On the other hand Arnold diffusion, which, as already mentioned, can
drive some orbits with initial conditions near the equilibrium point $L_4$ to
regions of the phase space far away from it, appears only in the spatial case
and not in the planar one. 

So in the present paper we study the spatial case, following a similar
procedure  to the one used in Paper I, in order to find out to what
degree the presence of Arnold diffusion changes the effective stability region.
Apart from the  fact that we consider the three dimensional case and not its
restriction on the plane, some minor changes of the scheme used in Paper I
improve slightly the estimation of the size of the stability region. In
particular the expansion of the Hamiltonian of the system in a power series
suitable for the application of the normal form scheme is computed with
higher accuracy than before. Also a more accurate calculation of the time
needed for an orbit to leave a particular region of the phase space around the
point $L_4$ (escape time) is provided. We were also able to compute the normal
form  to higher orders than in previous studies, both in the 3D and the 2D
case. In particular in the 3D case  we  have numerically
computed the normal form up to order 29, which is  a hard task,
since one has to manipulate functions with a much larger number of
coefficients, than in the case of order 22 (Celletti \& Giorgilli
\cite{celgior}). In the 2D case we computed the normal form up to order 49
(instead of order 34 in Paper I).

The paper is organized as follows. In Sect. 2 we present the Hamiltonian we
use, referring to all the canonical transformations needed to bring it in a
form suitable for the application of the normal form scheme. Also in this
section we sketch the procedure of computing the normal form. The main results
of the paper, concerning the estimation of the size of the regions of
effective stability both in the spatial (3D) and the planar (2D) case, are
presented in Sect. 3. The application of the above results to real asteroids
is also included in Sect. 3. In Sect. 4 we discuss some issues concerning
the effectiveness of our scheme. Finally in Sect. 5 we summarize our
results. 

%________________________________________________________________

\section{The Hamiltonian and the normal form of the system}

The spatial restricted problem of three bodies, in particular for the Sun
(S), Jupiter (J) and asteroid (A) system can be described as follows: we study
the motion of an asteroid A of infinitesimal mass, orbiting in the
gravitational field of two primaries S and J with masses equal to $1-\mu$ and
$\mu$ respectively, which are assumed to revolve in circular orbits around
their common center of mass. 

We introduce a uniformly rotating
frame (O, $q_1$, $q_2$, $q_3$) so that its origin is located at the center of
mass of the Sun-Jupiter system, with the Sun always at the point ($\mu$, 0, 0)
and Jupiter at the point ($1-\mu$, 0, 0). The physical units are chosen so
that the distance between Jupiter and the Sun is 1, $\mu=9.5387536 \cdot
10^{-4}$  and the angular velocity of Jupiter is 1. The time unit is
$(2\pi)^{-1}\,T_{J}$, where $T_{J}$ is the period of the circular motion of
Jupiter around the Sun. So the age of the universe is about $10^{10}$ time
units. The Hamiltonian of the system is :   
\begin{eqnarray}
H &=&\frac{1}{2} (p_{1}^{2} + p_{2}^{2}+ p_{3}^{2}) + q_{2}p_{1} - q_{1}p_{2}
               \nonumber \\
  & & -\frac{1-\mu}{\sqrt{(q_{1}-\mu)^{2}+q_{2}^{2}+q_{3}^{2}}} 
                \nonumber \\
  & & -\frac{\mu}{\sqrt{(q_{1}+1-\mu)^{2}+q_{2}^{2}+q_{3}^{2}}} \; .
    \label{eq:ham1}
\end{eqnarray}
The coordinates of the Lagrangian point $L_4$ are:
$q_1=\mu-\frac{1}{2}$,  $q_2=\frac{\sqrt{3}}{2}$, $q_3=0$, $p_1=-\frac{\sqrt{3}}{2}$,
$p_2=\mu-\frac{1}{2}$, $p_3=0$.

In order to bring the Hamiltonian to a form suitable for the application of
the normal form scheme we perform a sequence of transformations. The first
step is to introduce a uniformly rotating frame with its origin on the Sun (S)
using the generating function 
\[
W_3 = - (Q_1 + \mu) p_1 - Q_2 p_2 - Q_3 p_3 + \mu Q_2 \; ,
\]
where  $Q_1$, $Q_2$, $Q_3$, $P_1$, $P_2$,  $P_3$ are the heliocentric
coordinates. 

It is known that the projection on the plane of
Jupiter's orbit, of the stability region around $L_4$,  is a banana--shaped
region which lies close to the circle with center the Sun and radius equal to
the Sun-Jupiter distance. Since the plane of Jupiter's orbit is a symmetry
plane for the system, a good choice for describing this region are the
cylindrical coordinates $P$, $\Theta$, $Z$, which are introduced by the
generating function  \[
W_3 = - P(P_1\, cos\Theta + P_2 \,sin\Theta) -Z P_3 \; . 
\]
In this system of coordinates $L_4$ is located at $P=1$,
$\Theta=\frac{2\pi}{3}$, $Z=0$, $p_P=0$, $p_{\Theta}=1$, $p_Z=0$.

We move the origin of the coordinate system to the point $L_4$ using the
generating function
\[
W_2= p_x (P-1) + (p_y+1)\Theta - \frac{2\pi p_y}{3} + p_z Z \; .
\]
Then the Hamiltonian becomes:
\begin{eqnarray}
H &=&\frac{1}{2} \left[p_{x}^{2} + \frac{(p_{y}+1)^{2}}{(x+1)^{2}}+
        p_{z}^{2}\right] -p_y                \nonumber \\
  & & -\mu(x+1)\cos\left(y+\frac{2\pi}{3}\right) -
       \frac{1-\mu}{\sqrt{(x+1)^2+z^2}}             \nonumber \\
  & &  -\frac{\mu}{\sqrt{(x+1)^2+z^2+1+2(x+1) \cos\left(y+\frac{2\pi}{3}\right)}}     
     \; .
\label{eq:ham2}  
\end{eqnarray}

We expand the above Hamiltonian  in Taylor series around the
point $L_4$ $(x = y = z = p_x = p_y = p_z = 0)$ using the computer algebra
platform ``Mathematica" (Wolfram Research Inc.). The program allows us to
compute the coefficients of the expansion using arbritrary precision
arithmetics, while in Paper~I the corresponding expansion was made with less
accuracy. This change improves the credibility of our computation. After the
Taylor expansion the Hamiltonian becomes \mbox{$H=\sum_{s=2} H_s$} where $H_s$
is a homogeneous polynomial of degree $s$ in $x, y,  z,  p_x,  p_y,  p_z$. The
second order term is \begin{eqnarray} H_{2}&=& \frac{1}{2}
(p_{x}^{2}+p_{y}^{2})-2xp_{y}+(\frac{1}{2}+          
\frac{9\mu}{8})x^{2}-\frac{9\mu}{8}y^{2}+            \nonumber \\
     & & \frac{3\sqrt{3} \, \mu}{4}xy + \frac{1}{2} (p_z^2+z^2) \; .
\label{eq:h2}  
\end{eqnarray}

The last change of variables 
\begin{equation}
(x, y,  z,  p_x,  p_y, p_z) \rightarrow (x_1,x_2,x_3,y_1,y_2,y_3)
\label{eq:trans}
\end{equation}
is performed in order to bring the quadratic part $H_2$ (Eq.~\ref{eq:h2})  
to the diagonal form
\begin{equation}
H_2=\frac{1}{2}\sum_{j=1}^{3} \omega_j (x_j^2+y_j^2) \; .
\label{eq:quadr}
\end{equation}
Since the term $\frac{1}{2}(p_z^2+z^2)$ in Eq.~(\ref{eq:h2}) has this form with
$\omega_3=1$, the canonical transformation (Eq.~\ref{eq:trans}) must bring to
diagonal form the part of $H_2$ which corresponds to the planar case and
depends only on $x$, $y$, $p_x$, $p_y$. This transformation is the one
performed in Paper I, but we include it to make the paper self-consistent. So
we have  \begin{equation}
(x, y, p_x, p_y)^T=C\, (x_1,x_2,y_1,y_2)^T \; ,
\label{eq:trans2}
\end{equation}
where $T$ denotes the transpose matrix and C is the matrix
\begin{equation}
C=(e_1 m_1^{-\frac{1}{2}}, e_2 m_2^{-\frac{1}{2}},f_1 m_1^{-\frac{1}{2}}, f_2
        m_2^{-\frac{1}{2}})
\label{eq:matrix}
\end{equation}
with
\[
e_j=\left( \frac{8\omega_{j}^{2}+4\sqrt{3} \alpha+9}{8}, 
      \frac{4\alpha+3\sqrt{3}}{8}, 0,
      \frac{4\sqrt{3}\alpha+9}{4} \right)^T  \; ,       
\]
\[
f_j=\left( 
     0, 2\omega_j, 
     \frac{\omega_{j}(8\omega_{j}^{2}+4\sqrt{3} \alpha+9)}{8},
     \frac{\omega_{j}(4\alpha+3\sqrt{3})}{8} 
     \right)^T   \; ,      
\]
\begin{eqnarray}
\frac{m_j}{\omega_j}&=&  
    \left(\frac{8\omega_{j}^{2}+4\sqrt{3} \alpha+9}{8}\right)^{2}- 
     2\left(\sqrt{3}\alpha+\frac{9}{4}\right)
     \nonumber  \\
    & &+\left(\frac{4\alpha+3\sqrt{3}}{8}\right)^{2} \; , 
     \nonumber
\end{eqnarray}
for j=1,2 and
\[
\omega_{1}=\sqrt{  \frac{1}{2}+\frac{1}{2} 
  \sqrt{1-\frac{27}{4}+4\alpha^{2}}} \; ,
\]
\[
\omega_{2}=-\sqrt{  \frac{1}{2}-\frac{1}{2} 
  \sqrt{1-\frac{27}{4}+4\alpha^{2}}} \; ,
\]
\[
\alpha=- \frac{(1-2\mu) 3 \sqrt{3} }{4} \; .
\]

We remark that the transformation of Eq.~(\ref{eq:trans2}) involves only
the variables $x$, $y$, $p_x$, $p_y$ since $z$, $p_z$ remain unchanged. For
notational consistency we complete the transformation of Eq.~(\ref{eq:trans2})
by putting $x_3=z$ and $y_3=p_z$.

All the above transformations were performed in order to bring the Hamiltonian
to the form
\begin{equation}
H(x_1,x_2,x_3,y_1,y_2,y_3) \, = \, \sum_{s\geq2} H_s (x_1,x_2,x_3,y_1,y_2,y_3)
 \; ,
\label{eq:hamold}
\end{equation}
where $H_s$ is a homogeneous polynomial of degree $s$ in the hereafter called
`old variables' $x_1$, $x_2$, $x_3$, $y_1$, $y_2$, $y_3$. The quadratic part
$H_2$ which is given by Eq.~(\ref{eq:quadr}) is the Hamiltonian of a system of
three harmonic oscillators with frequencies $\omega_1\simeq 9.967575\cdot
10^{-1}$,  $\omega_2 \simeq -8.046388\cdot 10^{-2}$ and $\omega_3=1$. The form
 of Eq.~(\ref{eq:hamold}) is suitable for the direct application of the
normal form theory, as it is described in detail by Giorgilli et al.
(\cite{gioretal}). We give a brief sketch of this procedure: we construct a
generating function $X$, of the form \begin{equation}
X \, = \, \sum_{s\geq3} X_s \; , 
\label{eq:gener}
\end{equation}
where $X_s$ is a homogeneous polynomial of degree $s$, so that the
corresponding canonical transformation brings the Hamiltonian to normal form
\begin{equation}
Z \, = \, \sum_{s\geq2} Z_s \; ,
\label{eq:normalform}
\end{equation}
where $Z_s$ is a homogeneous polynomial of degree $s$ in the new `normal
variables' $x'_1$, $x'_2$, $x'_3$, $y'_1$, $y'_2$, $y'_3$. The term ``normal
form" means, that $Z$ is a function of the quantities $I'_i=\frac{1}{2}
(x_i^{'2}+y_i^{'2})$ with $i=1,2,3$, so that the system is  formally
integrable (Birkhoff \cite{birk}; Gustavson \cite{gust}).

The generating function $X$ and the normal form $Z$ are computed by solving the
equation
\begin{equation}
T_X\,Z \, = \, H \; ,
\label{eq:equation}
\end{equation}
where $T_X$ is an operator whose action is defined as follows
\[
T_X\,Z= \sum_{k\geq1} F_k \; ,
\]
where
\[
F_k=\sum_{s=1}^{k} Z_{s,k-s} \; ,
\]
\[
Z_{s,0}=Z_s \;,\; Z_{s,k}=\sum_{n=1}^{k} \frac{n}{k}\cdot L_{X_{2+n}}Z_{s,k-n}
\]
and $L_{X_k} Z_s=[X_k,Z_s] $, with $[\; , \; ]$ denoting the Poisson
bracket. The operator $T_X$ is linear, invertible and preserves products and
Poisson brackets (Giorgilli \& Galgani \cite{giorgal}). The 
 computer program that solves Eq.~(\ref{eq:equation})
and determines the generating function $X$ and the normal form $Z$ is
described by Giorgilli (\cite{gior79}).

Since the Hamiltonian in the old variables (Eq.~\ref{eq:hamold}) is an
infinite series, in practice we stop the expansion at some order $\tilde{r}$
and use the truncated Hamiltonian
\begin{equation}
H^{(\tilde{r})}=H_2+ H_3+ \cdots + H_{\tilde{r}} \; .
\label{eq:hamtr}
\end{equation}
Then for any fixed integer r with $3\leq r < \tilde{r}$ we solve
Eq.~(\ref{eq:equation})  defining a truncated generating function $X^{(r)}$
up to order r \begin{equation}
X^{(r)}=X_3+ X_4+ \cdots + X_r 
\label{eq:gentr}
\end{equation}
and constructing the  normal form $Z^{(r)}$ up to this order
\begin{equation}
Z^{(r)}=Z_2+ Z_3+ \cdots + Z_r + Y^{(r)} \; ,
\label{eq:nortr}
\end{equation}
where $Y^{(r)}$ is a remainder, actually a power series starting with terms of
degree r+1. So we have the equation
\begin{equation}
T_{X^{(r)}}^{-1}   H^{(r)} = 
\underbrace{Z_2+ Z_3+ \cdots + Z_r}_{normal \; form} +
\underbrace{ Y^{(r)}}_{remainder}    \; .
\label{eq:trequation}
\end{equation}
The first  term of the remainder $Y_{r+1}^{(r)}$ is also computed,
since it is needed for the estimation of the size of the effective stability
region as we will explain in Sect. 3. 

In Paper I, where the  planar problem (2D case) was studied, the power
series of 4 variables were truncated at order $\tilde{r}=35$. A function of 4
variables expanded up to order 35 requires 82,251 coefficients, while the
process of constructing the normal form requires the computation of several
functions with a total of 2,549,782 coefficients. In the spatial problem (3D
case) we use expansions of functions of 6 variables up to
order $\tilde{r}=30$. This is a much harder task compared to the 2D case
since a function of 6 variables expanded up to order 30 requires 1,947,792
coefficients and the program which calculates the normal form manipulates
55,929,459 coefficients.

%________________________________________________________________

\section{Estimation of the effective stability region}

%-------------------------------------
\subsection{Theoretical framework}

The transformed Hamiltonian $Z^{(r)}$ admits three approximate first 
integrals of the form 
\begin{equation}
I'_j(x',y')=\frac{1}{2}\left({x'_j}^2+{y'_j}^2\right)\; ,\;\;\; j=1,2,3 \; .
\label{eq:int} 
\end{equation}
The variation of these quantities in time, is given by 
\begin{equation}
\dot I'_j = [I'_j ,Z^{(r)} ] = [ I'_j ,Y^{(r)}] \; ,\;\;\;
j=1,2,3  \; ,
\label{eq:idot} 
\end{equation}
which is a power series starting with terms of degree $r+1$.

We introduce now suitable domains in the phase space, where we study the
stability properties of the system and also a norm in these domains, in order
to estimate the time variations of the three approximate integrals $\dot
I'_j$. 

For arbitrary fixed  positive constants $R_1$, $R_2$, $R_3$ we
consider a family of domains of the form
\begin{equation}
\Delta_{\rho R} = \left\{(x',y')\in {\bf R}^6\>:\> x_j^{'2}+y_j^{'2} \le  
     \rho^2 R_j^2 \right\} \; ,
\label{eq:domain} 
\end{equation}
where $\rho$ is a positive parameter and $x'$, $y'$  stand for $x'_1$, $x'_2$,
$x'_3$ and $y'_1$, $y'_2$,  $y'_3$ respectively. From the definition of the
domains $\Delta_{\rho R}$ it is evident that 
\begin{equation}
(x',y')\in \Delta_{\rho R}  \Rightarrow I'_j \leq  \frac{1}{2} \rho^2 R_j^2
\; ,\;\;\; j=1,2,3 \; .  
\label{eq:integranin} 
\end{equation}

The norm $\|f \|_{\rho R}$ of a homogeneous polynomial $f(x',y')$ of degree
$s$ in the domain $\Delta_{\rho R}$ does not exceed the quantity 
\begin{eqnarray}
\lefteqn{\|f\|_{\rho R} \leq } \nonumber \\
& &
\frac{\rho^s}{2^{s/2}}  
    \sum_{j_1j_2j_3k_1k_2k_3} \left|C_{j_1j_2j_3k_1k_2k_3}\right|  
    R_1^{j_1+k_1} R_2^{j_2+k_2} R_3^{j_3+k_3}  
\label{eq:norm} 
\end{eqnarray}
as shown in Paper I. $C_{j_1j_2j_3k_1k_2k_3}$ are the complex coefficients of
$f(x',y')$ when $f$ is transformed in complex variables $\xi$, $\eta$ via the
transformation $x'_j= (\xi_j+ i \eta_j)/\sqrt{2}$ , $y'_j= i(\xi_j- i
\eta_j)/\sqrt{2}$ for $j=1,2,3$. We remark that for the above norm the
elementary property \begin{equation}
\|f\|_{\rho R}= \rho^s\|f\|_{R} 
\label{eq:normineq} 
\end{equation}
holds.

We assume that the initial point of an orbit lies in the domain $\Delta_{\rho_0
R}$ for some positive value $\rho_0$, and we require that the orbit should be
confined inside a domain $\Delta_{\rho R}$ with $\rho > \rho_0$ for a finite
time interval. We shall refer to this time interval as the escape time $\tau$.
Since $\dot I'_j = dI'_j / dt$, we get 
\begin{equation}
dt \geq \frac{dI'_j}{\sup_{\Delta_{\rho R}} | \dot I'_j |} \; ,\;\;\; j=1,2,3 
 \; ,
\label{eq:dt} 
\end{equation}
where $\sup_{\Delta_{\rho R}} | \dot I'_j |$ denotes the maximum absolute
value of $ \dot I'_j$, or in other words the supremum norm of  $ \dot I'_j$,
over  the domain $\Delta_{\rho R}$. 

Giorgilli et al. (\cite{gioretal}) proved that the power series of the
remainder $Y^{(r)}$ is absolutely convergent in a domain  $\Delta_{\rho R}$
provided $\rho$ is small enough. So, assuming that $\rho$ is smaller than 
half of the convergence radius of the remainder $Y^{(r)}$ we can use the
approximate estimation  \begin{equation}
\sup_{\Delta_{\rho R}} | \dot I'_j | < 2 \| [ I'_j, Y_{r+1}^{(r)} \|_{\rho R}
  \stackrel{(\ref{eq:normineq})}{=} 2 \rho^{r+1} \| [ I'_j,
Y_{r+1}^{(r)} \|_R   \; ,
\label{eq:supi} 
\end{equation}
where $Y_{r+1}^{(r)}$ is the first term of the remainder. The term 
$Y_{r+1}^{(r)}$ is a homogeneous polynomial of degree $r+1$ and it is easily
computed as a byproduct of the program that calculates the normal form. The
validity of the above assumption is discussed in Sect. 4.

By integrating both parts of Eq.~(\ref{eq:dt}) and using Eq.~(\ref{eq:supi}), 
we estimate the minimum escape time as 
\begin{eqnarray}
\lefteqn{\tau_r (\rho_0,\rho) =} \nonumber \\
& &\min_{j=1,2,3} 
   \frac{R_j^2}{2 (r-1) \| [I'_j, Y_{r+1}^{(r)} ]\|_R }
    \left[ \frac{1}{\rho_0^{r-1}} - \frac{1}{\rho^{r-1}} \right] \; .
\label{eq:minesc}
\end{eqnarray} 
The above quantity depends on the order $r$ up to which the normal form is
constructed and on the radii $\rho_0$ and $\rho$ of the initial and final
domain respectively. In Eq.~(\ref{eq:minesc}) we  keep the smallest value with
respect to $j$ ($j=1,2,3$), because when the orbit is outside the disk with
radius $\rho R_j$ in one of the planes $(x'_j,y'_j)$ it is also
outside $\Delta_{ \rho R}$.

In order to have the minimum escape time as a function of $\rho_0$ we 
eliminate the dependence of  $\tau_r (\rho_0,\rho)$ on
$\rho$ and $r$. It is evident that for a given order $r$ the escape time goes
to infinity  as the radius of the outer domain grows. So by fixing $\rho$ to
be equal to $\lambda \rho_0$, with $\lambda > 1$ we get
\begin{eqnarray}
\lefteqn{\tau_{r,\lambda} (\rho_0) =} \nonumber \\
& &     \min_{j=1,2,3} 
   \frac{R_j^2}{2 (r-1) \rho_0^{r-1} \| [I'_j, Y_{r+1}^{(r)} ]\|_R }
    \left[ 1 - \frac{1}{\lambda^{r-1}} \right] \; .
\label{eq:minesc2} 
\end{eqnarray}

By giving a fixed value to $\lambda$ we eliminate the dependence of the
minimum escape time on the radius of the final domain. In particular, we put
$\lambda=1.2$, which means that the radius of the final domain is $20 \%$
greater than the radius of the initial domain.

The next step is to optimize the minimum escape time with respect to $r$. For
$\lambda=1.2$ we compute  $\tau_{r,1.2}(\rho_0)$ via Eq.~(\ref{eq:minesc2}) 
for $r$ running from 3 to the maximum order $\tilde{r}-1$, for every value of
$\rho_0$. We choose the optimal order $r_{opt}$ of the expansion as the one
that gives the maximum value of the  escape time. Thus we get the maximum
escape time $T$ as function of only the radius $\rho_0$ of the initial domain:
\begin{equation} 
T(\rho_0) = \max_{3\leq r < \tilde{r}}  \tau_{r,1.2}(\rho_0) \; .
\label{eq:tmax} 
\end{equation}

The maximum order of the power expansions in the 3D case
is $\tilde{r}=30$, which means that the normal form was computed up to order
29 and the first term of the remainder is of order 30. So it becomes clear
that $r_{opt} \leq 29$.

   \begin{table*}
      \caption[]{Estimated stability region for some real asteroids. CN is the
                  catalog number of the asteroids and $R_1$, $R_2$, $R_3$
                  are the radii corresponding to the initial data. The radius
                  $\rho_0$ of the effective stability region is computed in
                  the spatial (3D) case where the normal form was constructed
                  up to order 29 and in the planar (2D) case where the normal 
                  form was constructed up to order 49. An asteroid is inside
                  the stability region if $\rho_0 \geq 1$. The optimal order
                  of the expansion $r_{opt}$ is also reported in both cases.
                  The table is sorted in decreasing order with respect to
                  the value of $\rho_0$ for the spatial case.}          
\[
         \begin{tabular}{llllllll}
           \hline
            \noalign{\smallskip}
     &     &     &     &  \multicolumn{2}{c}{3D case} &\multicolumn{2}{c}{2D
case} \\  \cline{5-6} \cline{7-8}
 CN & $R_1$ & $R_2$ & $R_3$ & $\rho_0$ & $r_{opt}$ & $\rho_0$ &
$r_{opt}$   \\        
           \noalign{\smallskip}
            \hline
            \noalign{\smallskip}
$89211605$&$0.033150$&$0.019594$&$0.013270$&$1.0105 $&$29$&$1.1722$&$35$ \\
$2357$&$0.042346$&$0.028509$&$0.049056$&$6.3957\cdot
10^{-1}$&$29$&$8.7950\cdot 10^{-1}$&$36$ \\ 
$88181612$&$0.031302$&$0.002101$&$0.068011$&$6.3294\cdot
10^{-1}$&$28$&$1.5364$&$33$ \\
$41790004$&$0.016517$&$0.031063$&$0.068685$&$6.2154\cdot
10^{-1}$&$29$&$1.1687$&$36$ \\ 
$5257$&$0.031836$&$0.042424$&$0.031626$&$6.1978\cdot
10^{-1}$&$29$&$8.0918\cdot 10^{-1}$&$38$ \\
$4867$&$0.233120$&$0.736264$&$0.468336$&$4.0589\cdot
10^{-2}$&$29$&$5.1663\cdot 10^{-2}$&$36$ \\
$1867$&$0.216926$&$0.758254$&$0.466126$&$3.9819\cdot
10^{-2}$&$29$&$5.0305\cdot 10^{-2}$&$36$ \\
$88172500$&$0.242971$&$0.902082$&$0.524644$&$3.3747\cdot
10^{-2}$&$29$&$4.2302\cdot 10^{-2}$&$36$ \\
$2363$&$0.293736$&$1.012524$&$0.553082$&$3.0026\cdot
10^{-2}$&$29$&$3.7667\cdot 10^{-2}$&$36$ \\
$1208$&$0.361925$&$0.997579$&$0.543939$&$2.9917\cdot
10^{-2}$&$29$&$3.7807\cdot 10^{-2}$&$36$  \\
            \noalign{\smallskip}
            \hline
         \end{tabular}
      \]
   \label{tab1}
   \end{table*}

   \begin{figure}
\resizebox{\hsize}{!}{\includegraphics{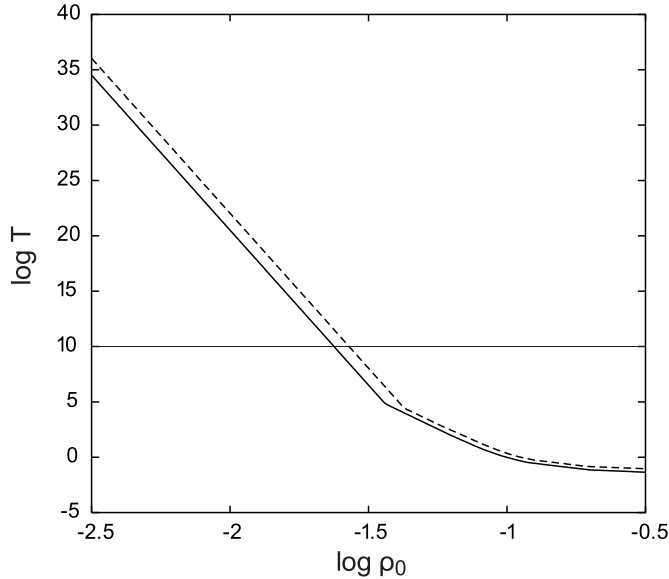}}
      \caption[]{
The logarithm of the maximum escape time $\log T$ as a function of the
logarithm of the radius of the initial domain $\log \rho_0$, in the spatial
3D case (solid line) and in the planar 2D case (dashed line). In both cases the
normal form has been computed up to order 29. The horizontal line marks the
time that corresponds to the age of the universe. }
         \label{Fig1}
   \end{figure}

%-------------------------------------
\subsection{General results}

For a general discussion and for making the results comparable to the ones
found in Paper I for the 2D case, we put $R_1=R_2=R_3=1$. In 
Fig.~\ref{Fig1} we plot the logarithm of the maximum escape time ($\log T$)
as a function of the logarithm of the radius of the initial domain ($\log
\rho_0$), in  the spatial (3D) and the planar (2D) cases. In both cases the
normal form is computed up to order 29. Taking an initially small
domain around the $L_4$ point, all the orbits are confined inside a slightly
larger domain (with $20\%$ greater radius), for very long time intervals,
while for large initial domains (large
values of $\rho_0$), the escape time is small.

A meaningful time interval for the system is the estimated age of the
universe, which is in our time units $10^{10}$. This value is marked in
Fig.~\ref{Fig1} by a horizontal line and it corresponds to $\log \rho_0
\simeq -1.625$, i.e. $\rho_0 \simeq 2.371 \cdot 10^{-2}$ in the 3D case and
to $\log \rho_0 \simeq -1.570$, i.e. $\rho_0 \simeq 2.692 \cdot 10^{-2}$ in
the 2D case. In both cases the optimal order of the expansion is $r_{opt}=29$.
The best previously  found estimation of the radius of the effective stability
region, was obtained in Paper I in the 2D case, namely $\rho_0 \simeq 2.911
\cdot 10^{-2}$, where the normal form was computed up to order 34. In that case
the optimal order was $r_{opt}=34$. The radius of the effective stability
region in the spatial case is about $12 \%$ smaller than the one computed for
the planar case for $r_{opt}=29$ and about $19 \%$ smaller than the one found
in Paper I for $r_{opt}=34$. So, the estimated stability region in the 3D case
is a realistic one since it is comparable to the region found in Paper I.

   \begin{figure}
\resizebox{\hsize}{!}{\includegraphics{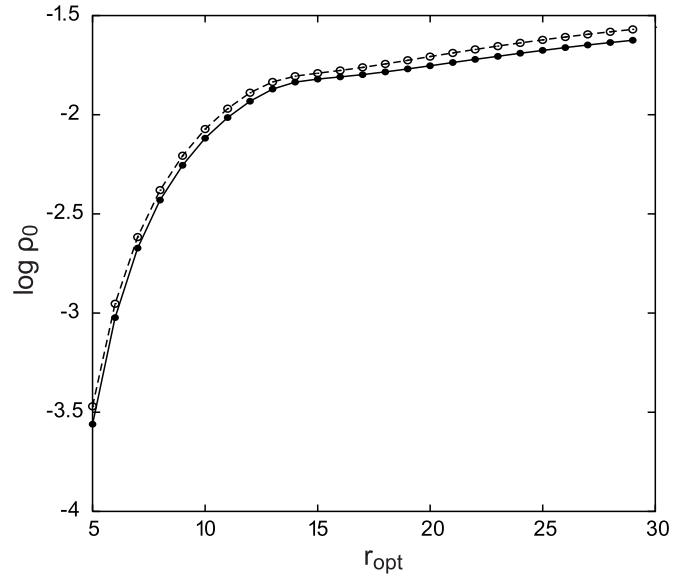}}
      \caption[]{
The logarithm of the radius $\log \rho_0$ of the effective stability
region which ensures stability for time equal to the age of the universe, as a
function of the optimal order $r_{opt}$ of the expansion of the normal form,
in the spatial 3D case (solid line) and in the planar 2D case (dashed line).
In both cases the normal form has been computed up to order 29.}
         \label{Fig2}
   \end{figure}

The estimated radii in the 3D and the 2D cases, for the same order of expansion
of the normal form, are relatively close to each other since they are of the
same order of magnitude, with the radius computed in the 3D case being always
smaller, as seen in Fig.~\ref{Fig2}. Thus the Arnold diffusion, which appears
only in the 3D case, does not affect the size of the effective stability
region significantly. 

We notice in Fig.~\ref{Fig2} that up
to order $r_{opt}=14$ the increment of the order $r_{opt}$ improves the
estimation of the radius significantly both in the 3D and the 2D case. For
orders greater than 15 the increment of the order leads to big increment of
the computational effort needed for the construction of the normal form
(mainly in the 3D case), but to relatively small improvements for the estimated
radii. For instance, the radius found in the 2D case for $r_{opt}=13$ is almost
equal to the one found for $r_{opt}=14$ in the 3D case ($\log \rho_0 \simeq
-1.84$) which was obtained by computing  almost 20 times more terms for the
various expansions, than in the 2D case. As $r_{opt}$ increases things become
even more difficult. For $r_{opt}=29$ in the 3D case we find  almost the same
radius as in the 2D case for $r_{opt}=25$  by computing almost
86 times more terms compared to the 2D case. So, it is evident that pushing
the computation of the normal form to higher orders becomes impractical for
the 3D case. On the other hand this can be done more easily in the 2D case.

According to Nekhoroshev theory, the series arising from the classical
perturbation theory have an asymptotic character in the sense  that 
 one should not exceed  an optimal value for the order of expansion, which
gives the best possible result. From Fig.~\ref{Fig2} we see that this limit
is greater than 30, both for the 3D and the 2D case. In
Paper I, where the construction of the normal form was done up to order 34 for
the 2D case, the limit $r_{opt}$ was not  reached.

By computing the normal form up to order 49 for the planar problem, so that the
first term of the remainder is of order 50, we find the maximum value of the 
optimal order of the expansion of the normal form to be $r_{opt}=38$, and the
corresponding value of $\log \rho_0$ equal to $-1.506$, i.e. $\rho_0 \simeq
3.119 \cdot 10^{-2}$. This value is about $7 \%$ greater than the one found in
Paper I for $r_{opt}=34$. This is the best estimation of the radius of the
stability region one can achieve using the particular theoretical framework, 
since expansions of the normal form to orders greater than 38 do not improve
the results.

%-------------------------------------
\subsection{Application to real asteroids}

In order to apply the above results to the real solar system we examine if 98
real asteroids, which are located near the Lagrangian point $L_4$, are inside
the estimated effective stability region. Using the same catalog as in Paper
I we extract the elements of the Trojan 
asteroids at the epoch,  December 14, 1994, J.D.$=2449700.5$. We also find the
elements of Jupiter at the same epoch. Assuming that the orbit of Jupiter is
circular we find the position of the $L_4$ point and compute the coordinates
$(Q_1,Q_2,Q_3,P_1,P_2,P_3)$ of the asteroids in the rotating heliocentric
system with the $z$ axis orthogonal to the plane of Jupiter's orbit. By using
the canonical transformations described in Sect.~2 we find the position of
every asteroid in the  $(x_1,x_2,x_3,y_1,y_2,y_3)$ coordinates which
diagonalize the quadratic part (Eq.~\ref{eq:h2}) of the Hamiltonian. Then we
define the radii $R_1$, $R_2$, $R_3$ for every asteroid as
$R_j=\sqrt{x_j^2+y_j^2}$ for $j=1,2,3$. Using these values we determine the
radius $\rho_0$ of the effective stability region in the way described
previously in this section. It is evident from the definition of $R_1$, $R_2$,
$R_3$, that an asteroid is inside the stability region if $\rho_0 \geq 1$. 

In the above procedure we use exactly the same asteroids and their elements
at the same epoch, as in Paper~I. In this way our results can be
compared directly to the ones obtained in Paper I where the effective
stability of 4 real asteroids was guaranteed. We apply this procedure both
to the spatial case, where the normal form is computed up to order 29 and to
the planar case,  where the normal form is computed up to order 49. We remark
that in order to find the radius $\rho_0$ in the 2D case only the values of
$R_1$ and $R_2$ are used, since we project the real asteroids on the plane of
Jupiter's orbit.

In Table  \ref{tab1} only the results for the 5 best and the 5 worst cases of
the radius $\rho_0$ in the 3D case are presented. For every asteroid  the
corresponding results in the 2D case are also reported. In the 3D case one real
asteroid is inside the stability region, while in the worst case the estimated
value of the stability region's radius is smaller by a factor 34. In all
cases the optimal order of the normal form is the maximum possible, 
$r_{opt}=29$, which means that the results may be improved for higher orders.

In the 2D  case the results are slightly better than the ones achieved in Paper
I, mainly because we performed the expansion of the normal form up to a higher
order. This improvement does not change the results significantly. Four real
asteroids (three reported in Table 1 and one not shown in that table) are
inside the planar stability region. In the worst case for the planar problem
(asteroid 2363) a factor 27 is needed in order for the asteroid to be safely
inside the stability region. The optimal order for all asteroids is $r_{opt}
\leq 38$, although the expansion of the normal form was performed up to order
49. So the computation of the normal form to orders higher than 38 does not
improve the estimations in the 2D case.

We remark that one would expect to find fewer asteroids inside the stability
region in the 3D case than in the 2D case,  since the spatial stability region
is projected on a plane in the 2D case. So 
points that are outside the spatial stability region may be projected inside
the planar stability region.

%________________________________________________________________

\section{Discussion}

In Sect. 3.1  we assumed that the supremum of $| \dot I'_j |$, which is an
infinite series, can be approximated by the norm of the first term multiplied
by 2 (Eq.~\ref{eq:supi}). The  practical reason for doing this
approximation is that we cannot compute the whole infinite series, while on
the other hand its first term can be obtained easily. The above assumption for
the norm is valid if the estimated value $\rho_0 \simeq 2.371 \cdot 10^{-2}$
in the spatial case, is smaller than the half of the convergence radius of the
remainder $Y^{(r)}$. But estimating the convergence radius of the remainder is
not possible since it requires the computation of higher orders of the series.

One can attempt an estimation of the convergence radius
following Paper I. Since we have computed  the expansion of the generating
function $X$ (Eq.~\ref{eq:gener}) up to order 30, we can evaluate its
convergence radius by fitting the norms $\| X_s \|_R$ for $3 \leq s \leq 30$ 
with a geometric sequence, i.e. we look for constants $a$ and $b$ such that
\begin{equation}
\| X_s \|_R \leq a b^{s-3} \; .
\label{eq:gs} 
\end{equation}
In order to satisfy this condition  it is sufficient to set 
\begin{equation}
a =\| X_3 \|_R \; \; \; , \; \; \;   
b = \max_{3 \leq s \leq 30} \left( \frac{\| X_s \|_R}{\| X_3 \|_R}
    \right)^{\frac{1}{s-3}} \; .
\label{eq:ab} 
\end{equation}
Giorgilli et al. (\cite{gioretal}) proved that if the generating function $X$
satisfies Eq.~(\ref{eq:gs}) then the coordinate transformation $x'=T_X x$,
$y'=T_X y$ is absolutely convergent in a domain $\Delta_{\rho R}$ provided
$\rho$ is small enough. The same was proven for the inverse coordinate
transformation and for the transformation of any other function such as the 
actions $I'_j$ (Eq.~\ref{eq:int}). So, we fit with a geometric sequence the
norms of the  coordinate transformations $T_X
x_1$, $T_X x_2$, $T_X x_3$, $T_X y_1$, $T_X y_2$, $T_X y_3$ and of the
transformations of the approximate integrals $I'_j$  to
the old variables $T_X^{-1} I'_1$, $T_X^{-1} I'_2$, $T_X^{-1} I'_3$. 
The worst case ($T_X^{-1} I'_2$) gives $b \simeq 18.665$, which corresponds to
a convergence radius $\rho \simeq 5.358 \cdot 10^{-2}$. Considering this value
as a good indicator of the true convergence radius, we conclude that the
estimated radius of the stability region 
is smaller than the convergence radius of the series, by a factor $2.2$, so it
is safely inside the convergence domain and the approximation used in
Eq.~(\ref{eq:supi})  holds. 

The theoretical framework we used proved to be very efficient, since we are
able to guarantee the effective stability of one and four real asteroids in the
spatial and planar problem, respectively. On the other hand we have reached the
limits of its effectiveness, since the above results cannot be improved
significantly.

Computing the expansion of the normal form to higher orders in the 3D case
would slightly improve the estimations, as we can see from the respective 
estimations performed in the 2D case, where the optimal order was reached for
$r_{opt}=38$.

The estimation of the minimum escape time $\tau_r (\rho_o,\rho)$ 
(Eq.~\ref{eq:minesc}) was improved compared to the one used in Paper I. This
improvement was done by taking into account the dependence of
$\sup_{\Delta_{\rho R}} | \dot I'_j | $ on the radius $\rho$ of the final
region (Eq.~\ref{eq:supi}), while in Paper I this
quantity had been overestimated by considering  $\sup_{\Delta_{\rho R}} | \dot
I'_j | $ constant inside the domain $\Delta_{\rho R}$ and equal to its value
at the edge of the domain. The above change forced us to fix  the radius
$\rho$ of the final domain. In our calculations we used $\lambda = \rho /
\rho_0 =1.2$ but this value does not influence strongly the results. This can
be easily understood since, the quantity $\left[ 1 - (1 / \lambda^{r-1}
)\right]$ in Eq.~(\ref{eq:minesc2}) is very close to 1 for $\lambda > 1.1$
and $r=29$ (which is the order of expansion up to which we compute the normal
form). The improved estimation of the escape time does not change the results
drastically. For instance, in the  2D case for $R_1=R_2=1$ with the
normal form being computed up to order 34, we got $\rho_0 \simeq 3.005 \cdot
10^{-2}$, which is about $3 \%$ greater than the value 
$\rho_0 \simeq 2.911 \cdot 10^{-2}$ obtained in Paper I for the same case.

The norms $\| [I'_j, Y_{r+1}^{(r)} ]\|_R $ for $j=1,2,3$ in
Eq.~(\ref{eq:minesc2})  were estimated using Eq.~(\ref{eq:norm}). Since
this estimation is purely analytic it is certainly pessimistic. For this
reason we made numerical calculations of the norms in order to determine
whether the overestimation of the norms influence strongly the estimation of
the radius of the stability region. We used two Fortran codes.  Algorithm
GLOBAL (Boender et al. \cite{num1}), which had the best performance in all
cases, was used to obtain the norms, and algorithm SIGMA (TOMS 667)
(Aluffi-Pentini et al. \cite{num2}, \cite{num3}) was used for verification
purposes. GLOBAL is a stochastic algorithm for finding the maximum of a
real-valued function. In stochastic methods for optimization, the probability
of finding the global maximum approaches unity as the sample size of the
random initial values increases. This algorithm utilizes a combination of
sampling, clustering, and local search; it terminates with a range of
confidence intervals on the value of the global maximum. SIGMA is a global
optimization algorithm, which implements a method founded on the numerical
solution of a Cauchy problem for a stochastic differential equation inspired
by statistical mechanics. A global maximum of the function is sought by
monitoring the values of the function along trajectories generated by a
suitable discretization of a first-order stochastic differential equation.

In all cases the numerically found maximum of $\| [I'_j, Y_{r+1}^{(r)} ]\|_R $,
$j=1,2,3$ was located at the edge of the domain $\Delta_R$. So by limiting our
computation on the edge we were able to evaluate the maximum with even greater
accuracy. The analytically estimated maximum was greater than the one computed
numerically by a factor 28 in the worst case. Using the numerically calculated
norms in Eq.~(\ref{eq:tmax}) we found $\rho_0 \simeq 2.667 \cdot 10^{-2}$. This
value is only $1.12$ times greater than the one computed by using the
analytically estimated norms. So the improvement of the size of the stability
region is negligible compared to the factor 34  needed for all the real
asteroids to be inside the stability region. Thus the estimations based on
the analytically found norms are reliable.

%________________________________________________________________

\section{Summary}

We studied the spatial and the planar circular restricted problem of three
bodies in the spirit of Nekhoroshev theory, considering in particular the
problem of the stability of the Trojan asteroids around the Lagrangian point
$L_4$ in the Sun--Jupiter--Asteroid model. By constructing the normal form
with terms up to order 29 in the spatial case and up to order 49 in the planar
case and using analytical methods and numerical approximations, we made
realistic estimations of the size of the effective stability region. Our
results can be summarized as follows:
   \begin{enumerate}
 \item The estimated size of the effective stability region in the spatial
case is big enough to  include 1 real asteroid. 
The region where the most remote asteroid is located (out of the 98
real  asteroids we checked), is larger by a factor 34 compared to the
estimated stability region. This result improves significantly older estimates
(Giorgilli et al. \cite{gioretal}; Celletti \& Giorgilli \cite{celgior}) where
no real asteroid was inside the stability region and a factor 3,000 was needed
for the most remote asteroid to be inside the stability region.
  \item The radii of the effective stability region in the general spatial and
planer cases are close to each other for the same order of expansion of the
normal form, with the radius computed for the spatial case being always
slightly smaller. Thus, Arnold diffusion does not affect the size of the
effective stability region significantly.
   \item By computing the normal form in the planar case, to higher order than
in Paper~I, the optimal order of the expansion of the normal form was
actually reached and it was found to be $r_{opt}=38$. Also the estimation of
the stability region's size was  slightly improved compared to Paper~I. Four
real asteroids are found to be inside the stability region, as it was found
also in Paper~I.
      \item Several improvements in the computational
procedure, compared to previous works, were introduced in the present paper,
namely: a) the computation of the normal form to higher orders; b) a more
accurate estimation of the minimum escape time $\tau_r (\rho_0,\rho)$
(Eq.~\ref{eq:minesc}); c) the numerical estimation of the norms  $\| [I'_j,
Y_{r+1}^{(r)} ]\|_R $  for $j=1,2,3$, which also verified the reliability of
the analytically calculated norms via Eq.~(\ref{eq:norm}) and d) the use of an
arbitrary precision computer algebra system for the expansion of the
Hamiltonian.       
       \item The theoretical framework we used  reached the
limits of its effectiveness by providing the best possible results in the 2D
case, and comparable results in the 3D case. In order to have a non negligible
improvement of the estimated size of the effective stability region, one has
to choose better coordinates than the cylindrical ones used in this
paper, in the sense that these coordinates should be more adapted to describe
the banana--shaped region of the actual stability region around $L_4$.    
   \end{enumerate}

%________________________________________________________________

\begin{acknowledgements}

We would like to thank  Prof. G. Contopoulos and Dr. P. Patsis for several
comments, which greatly improved the clarity of the manuscript. This research
was supported by the European Union in the framework of E$\Pi$ET  and
K$\Pi\Sigma$ 1994-1999 and by the Research Committee of the Academy of Athens.

\end{acknowledgements}

\end{document}